\documentstyle[twocolumn,floats,aps,epsf]{revtex}

\begin{document}

\twocolumn[\hsize\textwidth\columnwidth\hsize\csname@twocolumnfalse%
\endcsname]

{\bf \noindent Comment on ``Canonical and Mircocanonical Calculations for Fermi
  Systems''}

\thispagestyle{empty}

\narrowtext

\vspace{0.5cm}

In the context of nuclear physics Pratt\cite{Pratt} recently investigated
noninteracting Fermi systems described by the microcanonical and 
canonical ensemble. As will be shown his discussion of the
model of equally spaced levels contains a flaw and a statement
which is at least confusing.

In his Letter Pratt gives an alternative derivation of
an expression (Eq.\ (3) of Ref.\ \onlinecite{Pratt}) in which the
canonical partition function $Z_A(\beta)$ of a system with $A$
fermions at inverse temperature $\beta$ is expressed in
terms of the partition function with fewer particles 
$Z_{A-n}(\beta)$\cite{Borrmann1}. Starting with 
$Z_0(\beta) \equiv 1$, $Z_A(\beta)$ can thus be calculated
recursively.  This result also
leads to an expression in which the canonical occupation 
probability $\left< n_k \right>_{\rm c}$ for a 
one-particle state with 
energy $\epsilon_k$ is expressed as a sum over terms which 
contain $Z_{A-n}(\beta)$ and 
$\exp{(- n \beta \epsilon_k)}$\cite{Borrmann2,Pratt}. 
$Z_A(\beta)$ can be expressed as the (discrete) 
Laplace transform of the number of states $N_{A,E_r}$
of a system with $A$ particles and fixed energy $E_r$. 
Here the index $r$ indicates, that for a finite system the 
energy is discrete and given as the sum over $A$ one-particle 
energies $\epsilon_k$.
Furthermore the product 
$Z_A(\beta) \left< n_k \right>_{\rm c}$ is given as
the Laplace transform of $N_{A,E_r} \left< n_k \right>_{\rm
  mc}$ where $\left< n_k \right>_{\rm mc}$ is the microcanonical 
distribution function.
Starting with the expressions for the canonical ensemble and 
using inverse Laplace transformation it is thus possible to obtain 
recursive relations for $N_{A,E_r}$\cite{Borrmann2} and 
$\left< n_k \right>_{\rm mc}$. 
They are given in Eqs.\ (7) and (8) of Ref.\ \onlinecite{Pratt}. 

\begin{figure}[thb]
\begin{center}
\leavevmode
\epsfxsize6.5cm
\epsffile{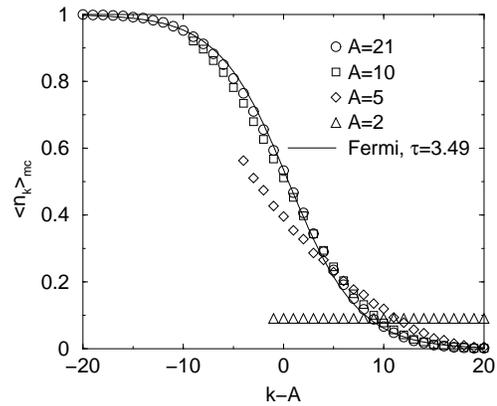}
\caption{The microcanonical occupation probability $\left< n_k
  \right>_{\rm mc}$ for excitation energy $20 g$ and different system
  sizes. For comparison the Fermi function for temperature $\tau
  \equiv k_B T/g = 3.49$ is shown as the solid line (see Ref.\ [1]).}
\label{fig1}
\end{center}
\end{figure}

As an example Pratt investigated a model with equally spaced 
(spacing $g$) nondegenerate fermionic levels.
The $A$-particle groundstate energy is given by 
$E_0(A)=g A(A+1)/2$ and the total energy $E_r(A)$ can be written as
$E_r(A) = E_0(A) + gr$, with $r=0,1,2, \ldots$. 
His discussion suggests that there is a direct connection between 
the number of states with an excitation energy $g r$ and the 
number of partitions of the integer $r$ into {\it distinct} parts. 
This is not the case.
Instead the model can be related to the {\it unrestricted partition
problem}\cite{KV,Arnaud}, i.e.\ the problem of finding the number of 
ways the integer $r$ can be written as a sum of smaller integers, 
where it is allowed to {\it repeatedly use the same integer.} 
The very fact that the same integer can be used more than once is
essential in connection with the mapping of the fermionic system on 
a bosonic one, a procedure which is known as 
{\it bosonization}\cite{KV}. 
For $r \leq A$, $N_{A,E_r}$ is given by the number of partitions
$p(r)$ of $r$ which is independent of $A$. 
For $r > A$ this mapping breaks down and 
$N_{A,E_r} < p(r)$\cite{KV,Arnaud}. In Fig.\ 3 of his Letter Pratt
presents data for $\left< n_k \right>_{\rm mc}$ and
$r=20$ calculated using his Eqs.\ (7) and (8). 
He does not give the number of particles for which the
data have been obtained. We have been informed by Pratt
that $r \leq A$, which implies that the data have to be
independent of $A$. For $r \leq A$ the microcanonical
occupation probability fulfills the {\it symmetry relation}
$\left< n_k \right>_{\rm mc} + \left< n_{A-k+1} \right>_{\rm mc}
=1$ and is a {\it monotonic} function of $k$\cite{KV,Arnaud}.
These observations are essential in connection with the question 
how the grand canonical distribution function 
(the Fermi function) is appoached in the thermodynamic limit.
Both requirements are apparently violated by the data presented
by Pratt. We have been informed by the author that in Fig.\ 3 
he accidentally used the wrong set of data. The data shown have been 
obtained for a model of randomly placed levels.  
Fig.\ \ref{fig1} shows the correct microcanonical
occupation probability for the equally spaced level model, 
$r=20$, and different system sizes. To
obtain the data for $r \leq A$ we have used the three different
algorithms presented in Refs.\ \cite{KV}, \cite{Arnaud},
and\cite{Pratt} which all lead
to the same curve. For $r > A$ the formulas presented in Refs.\
\cite{KV} and \cite{Arnaud} {\it cannot} be used and only the new 
result of Pratt which
does not make direct use of the mapping to the partition problem is
applicable. $\left< n_k \right>_{\rm mc}$ displays
monotonic behavior also for $r > A$. 
As a test of the validity of our data we have always checked the sum rule
$\sum_k \left< n_k \right>_{\rm mc}=A$ which is fulfilled up to very
high precision.

\noindent V.\ Meden and K.\ Sch\"onhammer\\
Institut f\"ur Theoretische Physik\\
Universit\"at G\"ottingen \\
Bunsenstr.\ 9 \\
D-37073 G\"ottingen \\
Germany

\end{document}